\documentclass[final,5p,times,twocolumn]{elsarticle}

\usepackage{amssymb}
\usepackage{lipsum}
\usepackage{amsfonts}
\usepackage{amsmath}
\usepackage{txfonts}
\usepackage{amsthm}
\usepackage{mathrsfs}
\usepackage{graphicx}
\usepackage{dcolumn}
\usepackage{multirow}
\usepackage{epstopdf}
\usepackage{url}  
\usepackage{natbib}
\usepackage{subfigure}
\usepackage{xcolor}

\newcommand{\jpsi}{$J/\psi$}

\journal{Physics Letters B}

\begin{document}

\begin{frontmatter}


\author{The STAR Collaboration}

\title{Measurement of inclusive \ensuremath{J/\psi} polarization in Ru+Ru and Zr+Zr\\ collisions at $\sqrt{s_{_{\rm NN}}} = 200$ GeV at STAR}

\begin{abstract}
The first measurement of inclusive \ensuremath{J/\psi} polarization at mid-rapidity ($|y^{J/\psi}| < 0.8$) in 200 GeV Ru+Ru and Zr+Zr collisions at $\sqrt{s_\mathrm{NN}} = 200$ GeV with the STAR experiment at RHIC is presented. \ensuremath{J/\psi} mesons are reconstructed through their di-electron ($e^+e^-$) decay channel. The polarization parameters ($\lambda _{\theta}$, $\lambda _{\phi}$) are measured as a function of the $J/\psi$ transverse momentum ($p_{\mathrm{T}}$) and collision centrality in both the helicity and the Collins-Soper frames. These polarization parameters are found to be consistent with zero across the measured \ensuremath{J/\psi} $p_{\mathrm{T}}$ range of $0.2<p_{\mathrm{T}}<10$ GeV/$c$ and across collision centralities within 0–80\% in both frames. These results are consistent with corresponding measurements $p$+$p$ collisions at the same collision energy and with transport-model calculations.
\end{abstract}

\end{frontmatter}



\section{Introduction}

The formation of the Quark-Gluon Plasma (QGP), a state of matter in which quarks and gluons are no longer confined within hadrons but exist freely, has been achieved through relativistic heavy-ion collisions at the Super Proton Synchrotron (SPS), the Relativistic Heavy Ion Collider (RHIC), and the Large Hadron Collider (LHC)~\cite{SPS_QGP_2000,Braun-Munzinger:2015hba,Harris:2023tti}. Among the probes used to study this deconfined state of matter, the \ensuremath{J/\psi} meson, a bound state of charm and anticharm quarks ($c\bar{c}$), plays a prominent role due to its sensitivity to the QGP-induced dissociation effect, referring to the breakup of \ensuremath{J/\psi} mesons in the QGP~\cite{jpsi_suppresion_in_QGP1986j,Rothkopf:2019ipj}. The reverse process, referred to as regeneration, i.e., the recombination of deconfined charm and anticharm quarks in the medium to form a \jpsi\ meson, could also occur~\cite{enhanced_jpsi_qgp_2001enhanced} and partially counterbalance the dissociation effect. 

Extensive measurements of nuclear modifications of \ensuremath{J/\psi} production yields in heavy-ion collisions, compared to those in $p$+$p$ collisions, have been carried out for studies of QGP properties~\cite{Tang:2020ame,Andronic:2025jbp}. Interpretation of these results is complicated by the various production mechanisms contributing to the inclusive \jpsi\ sample, including primordial and regenerated \jpsi. Primordial \jpsi\ mesons include prompt and non-prompt production, with the former referring to \ensuremath{J/\psi} produced directly in partonic scatterings and those from decays of excited charmonium states ($e.g.$, $\psi$(2S) and $\chi_{cJ}$), and the latter originating from decays of $b$-hadrons.
The presence of a QGP affects the prompt and non-prompt components differently, and they also have different dependences on collision energy and \jpsi\ kinematic variables. However, complete disentanglement has not been achieved and additional observables are needed.

Measurements of the \jpsi\ polarization~\cite{Theroyfaccioli2010towardsIntroduction} in heavy-ion collisions can potentially shed new light on QGP properties and distinguish different \jpsi\ production channels. For example, Ref.~\cite{TheroyBLioffe2003quarkonium} suggests that, in the presence of a quark–gluon plasma, non-perturbative effects responsible for quarkonium depolarization may be suppressed due to color screening, leading to a polarization of the surviving \jpsi\ closer to perturbative QCD expectations. On the other hand, \jpsi\ from $\psi(2S)$ decays should inherit the $\psi(2S)$ polarization\,\cite{psi2s_no_prolarization}, but those from $\chi_{cJ}$ decays can exhibit polarization patterns different from those of directly produced \jpsi\ since the photon emitted during decay is fully transversely polarized~\cite{photon_fully_polarization_2011determination}. While this is true for both heavy-ion and reference $p$+$p$ collisions, the relative contribution of $\chi_{cJ}$ decays to the inclusive \jpsi\ sample is expected to be altered in heavy-ion collisions due to the expected stronger suppression of $\chi_{cJ}$ production in the QGP compared to that of directly produced \jpsi~\cite{jpsi_suppresion_in_QGP1986j}. Since the feed-down contribution from $\chi_{cJ}$ states may exhibit different polarization characteristics than directly produced \jpsi\,, such a change in their relative fractions can consequently alter the observed inclusive \jpsi\ polarization in heavy-ion collisions with respect to $p$+$p$ collisions. Regenerated \jpsi\ are generally expected to exhibit small polarization due to the absence of initial spin alignment~\cite{Jpsi_Regeneration:2024}. However, possible polarization induced by medium effects such as vorticity or strong electromagnetic fields has also been discussed in the literature\,\cite{TheroyBaochi2024polarization,Chen2025VectorMesonSpinAlignment}. At RHIC energies, the regeneration contribution to inclusive \jpsi\ production is predicted to be modest because of the relatively small total $c\bar{c}$ production cross section, and is therefore not expected to dominate the observed polarization~\cite{TheroyBaochi2024polarization}.

\par \ensuremath{J/\psi} polarization in heavy-ion collisions has been measured at the LHC using Pb+Pb collisions at a center-of-mass energy per nucleon-nucleon pair ($\sqrt{s_{_{\rm NN}}}$) of 5.02~TeV in the \jpsi\ rapidity range of $2.5 < y < 4$~\cite{LHCHIC2021firstpolarizarion}. The polarization parameter $\lambda_{\theta}$ in the helicity (HX) frame (see definitions in Sec.~\ref{Methodology}) in Pb+Pb collisions differs from that in $p$+$p$ collisions at $\sqrt{s}=7$ TeV\,\cite{LHCLHCb2013polarizationpp} by $3.3\sigma$ within the $J/\psi$ transverse momentum ($p_{\rm T}$) range of $2-4$ GeV/$c$. There is a slight preference for positive $\lambda_{\theta}$ in Pb+Pb collisions, whereas the $p$+$p$ measurement shows a smaller negative value. Compared to LHC energies, the regeneration contribution to inclusive \ensuremath{J/\psi} production is expected to be smaller at RHIC due to the lower charm quark production cross section at lower collision energies. In addition, the fraction of non-prompt \ensuremath{J/\psi} originating from $b$-hadron decays is negligible at RHIC energies compared to the LHC, owing to the significantly lower bottom production cross section. Furthermore, the \ensuremath{J/\psi} polarization measurements at the LHC are typically performed at forward rapidity, whereas experiments at RHIC are positioned to measure it at mid-rapidity. These complementary kinematic regions provide an opportunity to improve our understanding of the behavior of \jpsi\ mesons in the QGP, given their complex production mechanisms. However, measurements of \jpsi\ polarization in heavy-ion collisions at RHIC have so far remained unavailable, primarily due to the low \ensuremath{J/\psi} production rate.

\par This letter presents the first measurement of inclusive \ensuremath{J/\psi} polarization at mid-rapidity ($|y^{J/\psi}|<0.8$) using the large samples of Ru+Ru and Zr+Zr collisions at $\sqrt{s_{\rm NN}}=200$ GeV by the STAR experiment. In Sec.~\ref{Methodology}, we introduce the polarization parameters used to quantify \ensuremath{J/\psi} polarization. In Sec.~\ref{Analysis details}, we explain how we perform electron identification, extract \ensuremath{J/\psi} yields, correct for detector acceptance and efficiency, and finally obtain the polarization parameters. The results are then presented in Sec.~\ref{Results and discussion}, followed by a summary in Sec.~\ref{summary}.

\section{Methodology}
\label{Methodology}
The degree of \ensuremath{J/\psi} polarization is reflected in the angular distribution of the decay products, which can be expressed as the following~\cite{Theroyfaccioli2010towardsIntroduction}:
\begin{equation}
 \begin{aligned}
    \emph W(\theta,\phi)\propto \frac{1}{3+\lambda_\theta}(1 &+\lambda_{\theta}\ \rm cos^2\theta+\lambda_{\phi}\ \rm sin^2\theta\ cos2\phi\\ &+\lambda_{\theta\phi}\ \rm sin2\theta\ cos\phi),   
     \end{aligned}
    \label{2D distribution}
\end{equation}
where $\lambda_{\theta}, \lambda_{\phi}, \lambda_{\theta \phi}$ are polarization parameters, and $\theta$ and $\phi$ are the polar and azimuthal angles of the positively charged daughter lepton in the \ensuremath{J/\psi} rest frame with respect to a chosen quantization axis ($z$-axis). This analysis involves the selection of two distinct reference systems\,\cite{Theroyfaccioli2010towardsIntroduction}:  the helicity (HX) frame and the Collins-Soper (CS) frame, both defined with respect to the \ensuremath{J/\psi} production plane. The production plane is spanned by the momenta of the colliding beams and the momentum of the \ensuremath{J/\psi}, with the $y$-axis being perpendicular to the production plane. The difference between the two frames lies in the definition of the $z$-axis. In the CS frame, the $z$-axis is defined as the bisector of the angle between one beam's direction and the opposite direction of the other beam in the \ensuremath{J/\psi} rest frame\,\cite{BookingParticlefaccioli2023polarization}. As a result, the CS frame is closely connected to the initial-state partonic kinematics. In the HX reference frame, the $z$-axis is determined by the \ensuremath{J/\psi} momentum direction in the center-of-mass frame of the collision, and therefore this frame is sensitive to polarization effects associated with the final-state hadronization process\cite{BookingParticlefaccioli2023polarization}.
The case where $(\lambda_{\theta}, \lambda_{\phi}, \lambda_{\theta \phi})$ equals $(1, 0, 0)$ corresponds to fully transverse polarization, while $(-1, 0, 0)$ indicates fully longitudinal polarization. The case of no polarization is represented by $(0, 0, 0)$\,\cite{Theroyfaccioli2010towardsIntroduction}. 

To extract the \ensuremath{J/\psi} polarization parameters, we integrate Eq.~\eqref{2D distribution} over $\phi$ and $\cos\theta$ respectively, yielding two one-dimensional (1D) distributions: 
       \begin{equation}
        \emph W(\rm cos\theta)=3\times\frac{1+\lambda_{\theta}\ \rm cos^2\theta}{2\times(3+\lambda_{\theta})},
        \label{Lambda_theta fucntion}
        \end{equation}
        \begin{equation}
        \emph W(\phi)= \frac{2\times\lambda_\phi}{(3+\lambda_\theta)\times2\pi}\rm cos2\phi.
        \label{Lambda_phi function}
        \end{equation}
The parameters $\lambda_{\theta}$ and $\lambda_{\phi}$ can therefore be obtained by simultaneously fitting the 1D angular distributions of daughter leptons using Eqs.~\eqref{Lambda_theta fucntion} and \eqref{Lambda_phi function}. 
     
While the measured polarization values depend on the selection of the quantization axis, one can construct a frame-invariant quantity ($\lambda_\mathrm{inv}$)\,\cite{Theroyfaccioli2010towardsIntroduction} to check the consistency of measurements in different frames. It is defined as
    \begin{equation}
    \lambda_\mathrm{inv}=\frac{\lambda_\theta+3\lambda_{\phi}}{1-\lambda_\phi}.
        \label{Lambda_inv}
    \end{equation}

\section{Analysis details}
\label{Analysis details}
    \subsection{Dataset, event and track selection}
        A sample of approximately $2\times10^{9}$ Ru+Ru and Zr+Zr collisions at $\sqrt{s_\mathrm{NN}}=200$ GeV is used for each system. The two systems combined due to their similar nuclear structure and energy density to enhance statistical precision. The minimum bias (MB) trigger, which requires a coincidence of signals from STAR's two Zero Degree Calorimeters (ZDCs) covering $|\eta|>6.3$~\cite{Xu:2016alq}, is employed to select events for this analysis. The primary subdetectors used in this analysis include the Time Projection Chamber (TPC), the Time-of-Flight (TOF) detector, and the Barrel Electromagnetic Calorimeter (BEMC). The TPC\,\cite{STARTPC2003star}, a gaseous drift detector with multi-wire proportional chambers and pad-row readout, can reconstruct trajectories of charged-particle tracks, measure their momenta, and provide information on ionization energy loss ($\mathrm{d}E/\mathrm{d}x$) for particle identification. It allows particle identification within $|\eta|<1$ across the full azimuth. Located outside the TPC, the TOF detector\,\cite{TOF2003single} measures a particle's flight time, which can be used to further distinguish electrons from hadrons. The TOF covers $|\eta|<0.9$ over the full azimuth. Between the TOF and the STAR magnet is the BEMC\,\cite{STARBEMC2003star}, a sampling calorimeter composed of layers of lead and plastic scintillator. It can be used to identify high-$p_{\rm T}$ electrons via their energy depositions in the BEMC within $|\eta|<1$ and the full azimuth. These subdetectors are housed within a solenoidal magnet that generates a uniform magnetic field along the beam direction with a strength of 0.5~T\,\cite{STARmagneticfield2003star}.

        \par The collision centrality is determined based on the measured  charged-particle multiplicity within $|\eta|<0.5$ and in comparison to a Monte Carlo Glauber simulation\,\cite{isobarCentrality2022search}.
        Event vertices are reconstructed from TPC tracks. Their positions along the beam axis ($v_{z}$) are required to lie within $-35~\mathrm{cm} < v_{z} < +25~\mathrm{cm}$ relative to the center of the TPC. The asymmetric $v_{z}$ cut is required because of the asymmetric vertex distribution due to on-line vertex selection. For \ensuremath{J/\psi} reconstruction, tracks originating from event vertices are used, and their $p_{\rm T}$ is required to be greater than or equal to 0.2 GeV/$c$, below which the TPC tracking efficiency drops sharply due to tight track curvatures and increased multiple scatterings in the material. The pseudorapidity of selected tracks is restricted to be within $|\eta| < 0.8$ to ensure uniform detector acceptance for different vertex positions. The Distance of Closest Approach (DCA) of tracks to the event vertex must be less than 1 cm to minimize contributions from secondary decays. The number of TPC hit points used to reconstruct the track (nHitsFit) should be at least 20 to ensure high momentum resolution, and the number of points used for calculating $\mathrm{d}E/\mathrm{d}x$ (nHitsDedx) should be no fewer than 15 to maintain good $\mathrm{d}E/\mathrm{d}x$ resolution. Additionally, the ratio of hit points on the track to the maximum possible number of hits along the track trajectory should exceed 0.52 to remove split tracks.
\subsection{Electron identification}
        The \ensuremath{J/\psi} meson is reconstructed through its di-electron decay channel. In the following discussion, ``electrons'' denotes both electrons and positrons unless otherwise specified. To identify electrons and reject hadrons, information from the TPC, TOF, and BEMC is used. The TPC provides particle identification capability through $\mathrm{d}E/\mathrm{d}x$ measurements. Specifically, the variable $n\sigma_{\mathrm{e}}$ is determined by quantifying the difference between the measured $\mathrm{d}E/\mathrm{d}x$ and the expected value for electrons based on the Bichsel function \cite{Bichselfunction2001comparison}, normalized by $R_{\mathrm{d}E/\mathrm{d}x}$, the resolution of $\ln(\mathrm{d}E/\mathrm{d}x)$. The variable is defined as follows:
\begin{equation}
n\sigma_{\mathrm{e}}=\frac{\ln(\mathrm{d}E/\mathrm{d}x)_{\mathrm{measured}}-\ln(\mathrm{d}E/\mathrm{d}x)_{\mathrm{theory}}^{\mathrm{e}}}{R_{\mathrm{d}E/\mathrm{d}x}},
    \label{nSigmaE}
\end{equation}
where $(\mathrm{d}E/\mathrm{d}x)_{\mathrm{measured}}$ is the measured energy loss and $(\mathrm{d}E/\mathrm{d}x)_{\mathrm{theory}}^{\mathrm{e}}$ is the theoretically calculated energy loss for electrons. Track momentum-dependent $n\sigma_{\mathrm{e}}$ cuts (Table~\ref{eIDCuts}) are applied to effectively suppress hadron contamination at low momentum, where the electron and pion $\mathrm{d}E/\mathrm{d}x$ bands overlap, while maximizing selection efficiency. For $0.2<p_{\rm T}<1$ GeV/$c$, TOF information is used to further improve the electron purity. Specifically, the selection $|1 - 1/\beta| < 0.025$ is applied, where $\beta$ is the particle speed, derived from time-of-flight measurements and normalized by the speed of light. 
\par For $p_{\rm T}>1$ GeV/$c$ and when the electron track is matched to a cluster in the BEMC, the BEMC information can be used to further distinguish between electrons and hadrons. The energy deposited in the BEMC for an electron is approximately equal to its momentum, whereas for hadrons, it is significantly lower.  Therefore, a requirement that$0.5<E_0/p<1.5$ is used to select electrons, where $E_0$ represents the highest tower energy in the matched BEMC cluster and $p$ is the track momentum. The electron identification criteria are summarized in Table \ref{eIDCuts}. It is worth noting that three matching scenarios are considered for electron identification in the range $1<p_{\rm T}<30$ GeV/$c$. If the track only matches a TOF hit (``only TOF''), electron identification is performed using $\beta$ and $n\sigma_{\rm e}$. If the track only matches a BEMC cluster (``only BEMC''), electron identification relies on $n\sigma_{\rm e}$ and $E_0/p$. For tracks that are matched to both TOF and the BEMC (``TOF \& BEMC''), electron identification uses $n\sigma_{\rm e}$, $\beta$, and $E_0/p$.
\begin{table}[htbb]
\centering

\caption{Electron identification cuts in Ru+Ru and Zr+Zr collisions at $\sqrt{s_{_{\rm NN}}} = 200$ GeV.}
\label{tab:systopo}
\resizebox{\linewidth}{!}{%
\begin{tabular}{cccc}
	\hline
 & $p_{\rm T}$ range& Selection criteria & Cuts\\
 \hline
 & \multirow{4}{*}{$0.2<p_{T} \leq 1$ GeV/$c$}&\multirow{2}{*}{$p \leq 0.8$ GeV/$c$}& $3\times p-3.15<n\sigma_{\rm e}<2$ \\
 & & &$|1/\beta-1|<0.025$ \\
  \cline{3-4}
 & & \multirow{2}{*}{$p> 0.8$ GeV/$c$} & $-0.75<n\sigma_{\rm e}<2$\\
 & & &$|1/\beta-1|<0.025$ \\
 \hline
 & \multirow{7}{*}{$1<p_{T}<30$ GeV/$c$} & \multirow{2}{*}{only TOF}&$-0.75<n\sigma_{\rm e}<2.$\\
 & & &$|1/\beta-1|<0.025$ \\
  \cline{3-4}
  &  & \multirow{2}{*}{only BEMC}&$-1<n\sigma_{\rm e}<2$\\
 & & &$0.5<E_{0}/p<1.5$ \\
 \cline{3-4}
   &  & \multirow{3}{*}{TOF \& BEMC}&$-1<n\sigma_{\rm e}<2$\\
 & & &$|1/\beta-1|<0.025$ \\
  & & &$0.5<E_{0}/p<1.5$ \\
	\hline
\end{tabular}
}
\label{eIDCuts}
\end{table}

    \subsection{Extraction of \ensuremath{J/\psi} yield} 
\begin{figure}[h!]
    \centering
    {\includegraphics[width=1\linewidth]{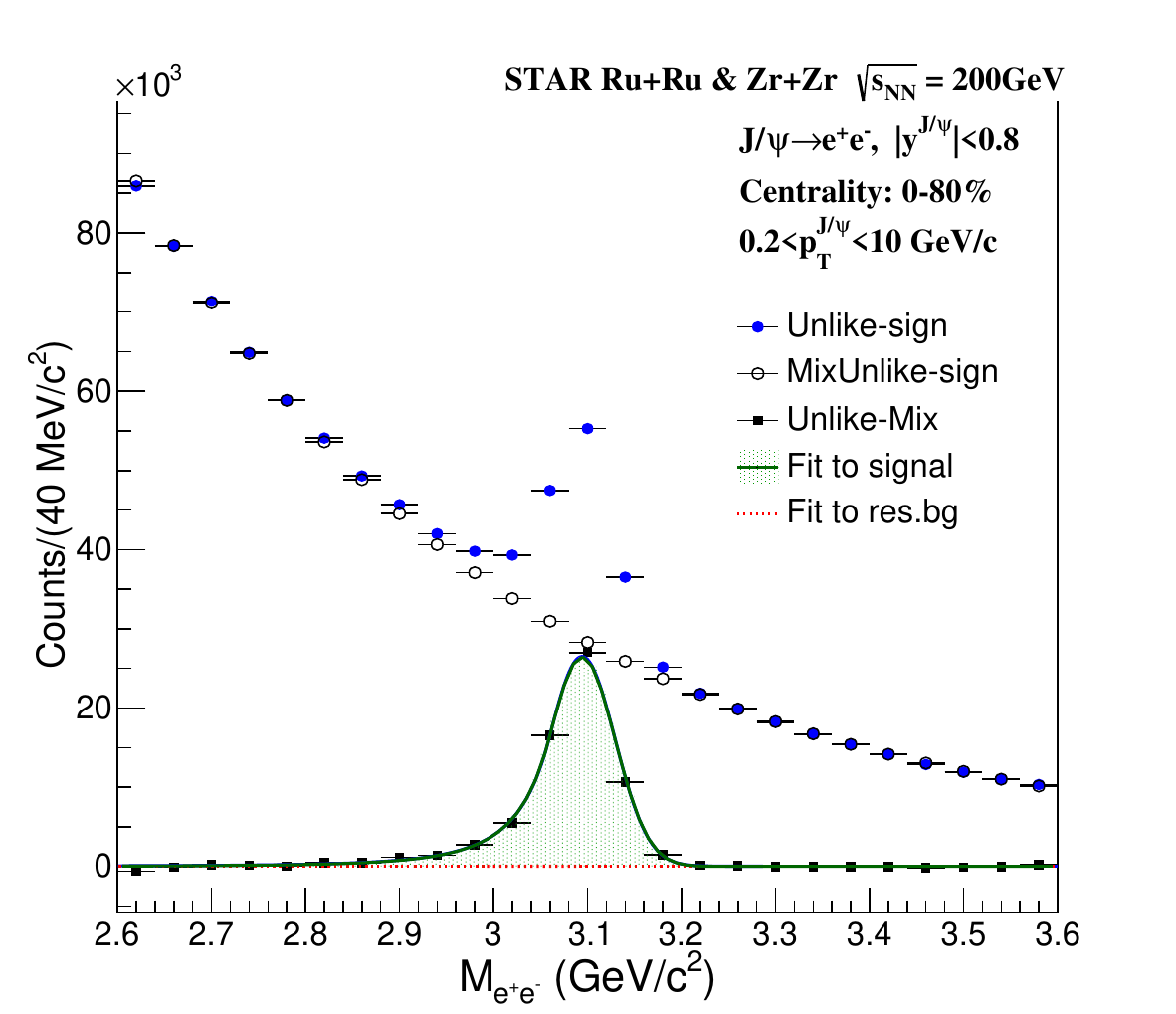}}
    \caption{Invariant mass distributions of di-electron pairs before (solid blue circles) and after (solid black squares) subtracting the combinatorial background estimated from event mixing (open black circles). The solid green curve with the hatched area represents a fit of the \ensuremath{J/\psi} invariant mass peak with a Crystal-Ball function, while the red curve denotes the fit to the residual background. The statistical uncertainties are smaller than the marker size.}
    \label{fig:Mass}
    \end{figure}

    The selected electrons and positrons ($e^{+}e^{-}$) are then paired to produce the invariant mass spectrum of \ensuremath{J/\psi} candidates within $0.2 < p^{J/\psi}_{\rm T} < 10$~GeV/$c$ and $|y^{J/\psi}| < 0.8$, as shown in Fig.~\ref{fig:Mass}, for the 0–80\% centrality class. To assess the background contribution, we perform a fit of the invariant mass distribution. The fitting procedure uses a Crystal-Ball function\,\cite{crystal_ball_1982study,crystal_ball_1983charmonium,crystal_ball_1986study,ROOT:RooCrystalBall} to characterize the \ensuremath{J/\psi} signal. The combinatorial background is constructed using the mixed-event technique~\cite{mix_event_STAR_phi_2004}, in which tracks from different events with similar global characteristics (such as collision centrality and primary vertex position) are combined to reproduce the uncorrelated background distribution. The mixed-event distribution is normalized to the same-event unlike-sign distribution in the invariant mass sideband region $3.3 < m_{ee} < 3.6 \mathrm{GeV}/ c^{2}$, outside the \ensuremath{J/\psi} signal window. A fourth-order polynomial function is used to account for the residual correlated background.

    All the parameters of the Crystal-Ball function, except the magnitude, are fixed according to a Monte Carlo (MC) simulation of the STAR detector performance with the simulated track momentum resolution tuned to match experimental data. To account for observed variations in the \ensuremath{J/\psi} mass shape, these parameters are fixed individually for each $\cos\theta$ or $\phi$ bin based on the simulation.
 
    \par 
    The raw \ensuremath{J/\psi} yield is determined by counting the entries within the mass window ($3 < M_{\rm e^+e^-} < 3.2$ GeV/$c^2$) and subtracting the combinatorial and residual backgrounds. This result is then corrected for the mass window efficiency, defined as the fraction of the fitted Crystal-Ball function integral contained within the chosen invariant mass window of $3.0 < M_{e^+e^-} < 3.2$ GeV/$c^{2}$. This correction yields a total of 67,682 $\pm$ 542 $J/\psi$ candidates. For the \ensuremath{J/\psi} polarization analysis, the sample is further divided into twenty bins of $\cos\theta$ ranging from -1 to 1, or fifteen bins of $\phi$ ranging from $-\pi$ to $\pi$, within each $p_{\rm T}^{J/\psi}$ or centrality interval. To ensure reliable signal extraction, \ensuremath{J/\psi} yields with a significance less than 3 in any bin are not considered in the subsequent analysis. The significance is defined as $S/\sqrt{S+B}$, where $S$ is the signal yield and $B$ is the background yield. The upper panels of Figs.~\ref{fig:EXtraction of polarization parameters HX} and \ref{fig:EXtraction of polarization parameters CS} display the raw \ensuremath{J/\psi} yield distributions as a function of $\cos\theta$ and $\phi$ for $0.2 < p_{\rm T}^{J/\psi} < 10$ GeV/$c$ and a centrality range of 0–80\% centrality in the HX and CS frames, respectively. 
    
    \subsection{Acceptance and efficiency}
    The different aspects of electron reconstruction and identification efficiencies, including TPC tracking efficiency, TOF and BEMC matching efficiency, and particle identification efficiency, are estimated either using a data-driven approach or based on a detector simulation. The TOF matching efficiency and electron identification efficiencies related to the 1/$\beta$ and n$\sigma_e$ selections are derived from analyzing a pure electron sample obtained from photon conversions in experimental data \cite{photonicConversiondetailed}. The 1/$\beta$ cut efficiency is estimated as the ratio between electrons matched to the TOF hits and those passing the $|1 - 1/\beta| < 0.025$ selection. A dependence of this efficiency on track $\eta$ is observed and taken into account. The n$\sigma_e$ cut efficiency is obtained via parameterizing n$\sigma_e$ distributions for electrons in narrow momentum bins with a Gaussian function and calculating the fraction of electrons falling into n$\sigma_e$ cut ranges based on the fit function. 
    On the other hand, TPC tracking, BEMC matching and BEMC-related electron identification efficiencies are evaluated through a detector simulation, where electrons are passed through a GEANT simulation of the STAR detector, embedded into real events and reconstructed in the same way as real data.
    
     \par The \ensuremath{J/\psi} acceptance and efficiency ($A$ $\times$ $\epsilon$) as a function of $\cos\theta$ or $\phi$ are determined using a toy Monte Carlo (ToyMC) simulation to fold in the electron efficiencies. The ToyMC simulation is set up as follows: the \ensuremath{J/\psi} azimuthal angle is uniformly distributed between $-\pi$ and $\pi$, while the \ensuremath{J/\psi} rapidity distribution is modeled based on the parametrization of the measurement in $p$+$p$ collisions at $\sqrt{s}=200$ GeV with a Gaussian function and restricted to $|y^{J/\psi}|<0.8$\,\cite{PHENIXRapidity2009j}. Daughter electrons from \ensuremath{J/\psi} decays are required to be within $|\eta|<0.8$ and weighted with their reconstruction and identification efficiencies.  Here, the TPC tracking, TOF matching, BEMC matching and BEMC-related identification efficiencies for electrons have been evaluated in bins of track $p_{\rm T}$, $\eta$ (divided into 40 bins from -1 to 1), and event vertex z position $v_{z}$ (divided into 6 bins from -35 to 25 cm). 
     \par The measured \ensuremath{J/\psi} $p_{\rm T}$ spectrum in Au+Au collisions at $\sqrt{s_{{\rm NN}}} = 200$ GeV~\cite{AuAu200_2019measurement} is used as an input to the ToyMC to determine the \ensuremath{J/\psi} $A$ $\times$ $\epsilon$. This is then used to correct the raw \ensuremath{J/\psi} yields extracted from 200 GeV Ru+Ru and Zr+Zr collisions, and the corrected spectrum is then used as the new input to the ToyMC. Since the angular distributions of the daughter electrons depend on the \ensuremath{J/\psi} polarization, which is not known {\it a priori}, an iterative procedure is employed. In the first iteration of the ToyMC simulation, \ensuremath{J/\psi} is assumed to be unpolarized and the resulting \ensuremath{J/\psi} $A$ $\times$ $\epsilon$ is used to correct the raw data and extract the \ensuremath{J/\psi} polarization parameters. These parameters are then fed into the ToyMC simulation for the next iteration. The process continues until the differences in the extracted polarization parameters between two consecutive iterations are 5 times smaller than their respective statistical uncertainties~\cite{STARLiuZhen2020polarization}. Typically, this convergence is achieved within 3-5 iterations, demonstrating the robustness and stability of the iterative method. 

    \subsection{Extraction of polarization parameters}
\begin{figure}
    \centering
    {\includegraphics[width=1\linewidth]{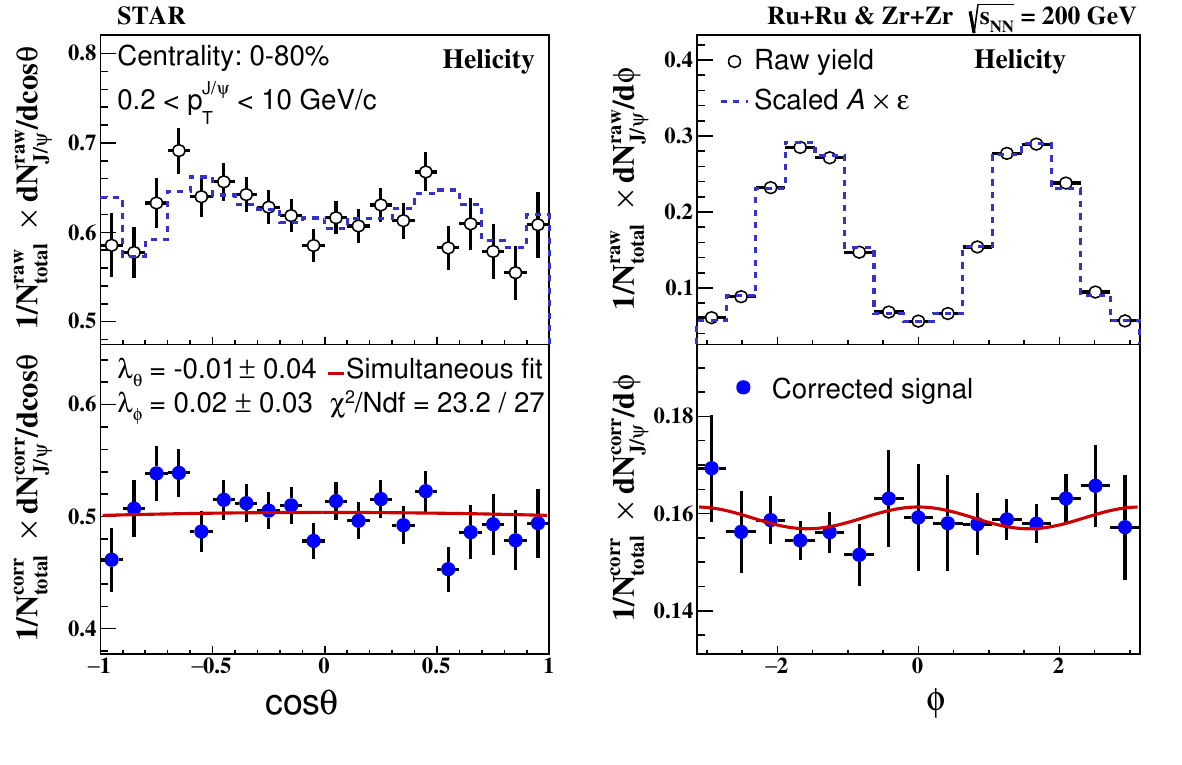}}
    \caption{Upper: raw \ensuremath{J/\psi} yield and its acceptance and efficiency ($A\times\epsilon$) from the final iteration as a function of daughter positron's $\cos\theta$ (left) and $\phi$ (right) in the helicity frame for $0.2<p_{\rm T}^{J/\psi}<10$~GeV/$c$, where $N_{\mathrm{total}}^{\mathrm{raw}}$ denotes the total raw \ensuremath{J/\psi} yield. Lower: \ensuremath{J/\psi} yields corrected for acceptance and efficiency, along with the simultaneous fit, where $N_{\mathrm{total}}^{\mathrm{corr}}$ denotes the total corrected \ensuremath{J/\psi} yield.}
    \label{fig:EXtraction of polarization parameters HX}
    \end{figure}

\begin{figure}
    \centering
    {\includegraphics[width=1\linewidth]{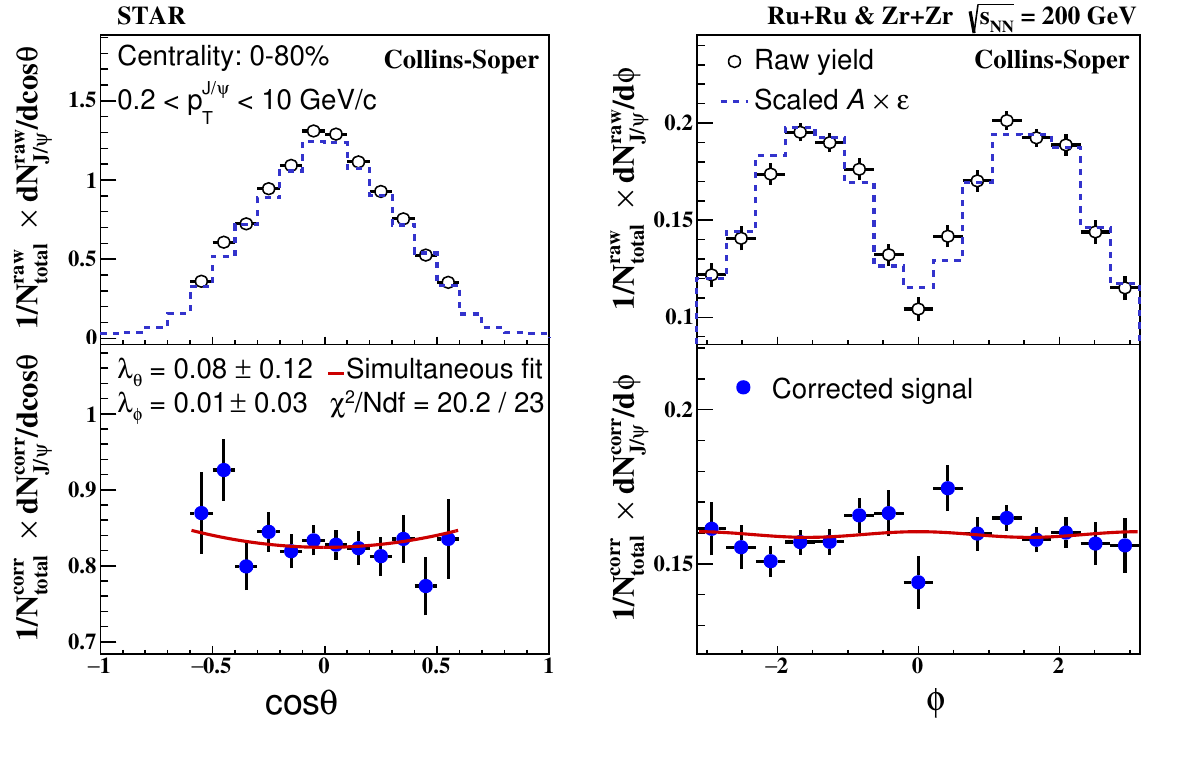}}
    \caption{Upper: raw \ensuremath{J/\psi} yield and its $A\times\epsilon$ from the final iteration as a function of daughter positron's  $\cos\theta$ (left) and $\phi$ (right) in the Collins-Soper frame for $0.2<p_{\rm T}^{J/\psi}<10$ GeV/$c$, where $N_{\mathrm{total}}^{\mathrm{raw}}$ is total $\rm{J}/\psi$ yield of raw data. Lower: \ensuremath{J/\psi} yields corrected for acceptance and efficiency, along with the simultaneous fit, where $N_{\mathrm{total}}^{\mathrm{corr}}$ denotes the total corrected \ensuremath{J/\psi} yield.}
    \label{fig:EXtraction of polarization parameters CS}
    \end{figure}

    Following the iterative procedure, the \ensuremath{J/\psi} efficiency multiplied by acceptance from the last iteration is shown in the upper panels of Fig.~\ref{fig:EXtraction of polarization parameters HX} as blue dashed curves. They are scaled to have the same integrals as the raw \ensuremath{J/\psi} yield distributions, shown as open circles in the same panels. The corrected \ensuremath{J/\psi} yields as a function of $\cos\theta$ (left) and $\phi$ (right), shown as filled circles in the lower panels, are obtained by dividing the raw yields by the corresponding acceptance and efficiency. The \ensuremath{J/\psi} polarization parameters ($\lambda_{\theta}$ and $\lambda_{\phi}$) can be extracted by simultaneously fitting the corrected yield distributions using Eqs.~\eqref{Lambda_theta fucntion} and \eqref{Lambda_phi function}.  The lower panels of Figs.\,\ref{fig:EXtraction of polarization parameters HX} and \ref{fig:EXtraction of polarization parameters CS} display the corrected \ensuremath{J/\psi} yield as a function of $\cos\theta$ and $\phi$ in the HX and CS frames, along with the simultaneous fit to both distributions, represented by red solid curves.
    
\subsection{Systematic uncertainties}
The following sources of systematic uncertainties are considered in this analysis: signal extraction, TPC tracking efficiency, and electron identification, all of which have comparable magnitudes. The total systematic uncertainties, determined by adding individual ones in quadrature, are listed in Tables~\ref{table_results pt} and \ref{table_results cent}.

\subsubsection{\ensuremath{J/\psi} signal extraction}
     The signal extraction systematic uncertainty is evaluated by varying different aspects of the extraction procedure: adjusting the fitting range from [2.6, 3.6] GeV/$c^2$ to [2.56, 3.64] GeV/$c^2$; changing the residual background function from a fourth-order polynomial to a third-order polynomial; modifying the normalization range for the mixed-event background from [3.3, 3.6] GeV/$c^2$ to [3.4, 3.6] GeV/$c^2$; switching the \ensuremath{J/\psi} yield extraction method from bin counting to fitting; and reducing the bin width of the invariant mass spectrum from 40 MeV/$c^2$ to 20 MeV/$c^2$. The Root Mean Square (RMS) of these variations is taken as the uncertainty. 
     
\subsubsection{Tracking efficiency}
To evaluate the uncertainty in the TPC tracking efficiency, track quality cuts are varied simultaneously in the analysis of both the real and simulated data. The specific variations include adjusting the DCA cut from $\text{DCA} < 1\text{ cm}$ 
to $\text{DCA} < 0.8\text{ cm}$ and $\text{DCA} < 1.5\text{ cm}$, changing 
$\text{nHitsFit} > 20$ to $\text{nHitsFit} > 15$, and changing 
$\text{nHitsDedx} > 15$ to $\text{nHitsDedx} > 10$. The Barlow method\,\cite{barlow2002systematic} is applied in assessing the changes in the extracted polarization parameters for each cut variation to suppress influences of statistical fluctuations, and the RMS of these changes is used as an estimate of the TPC tracking efficiency uncertainty.

\subsubsection{Electron identification}
The systematic uncertainty in electron identification arises from those in the TOF matching, n$\sigma_{e}$, 1/$\beta$, and BEMC efficiencies. For the TOF matching and $n\sigma_{e}$ cut efficiencies, the uncertainties 
are assessed by comparing results from different pure electron samples 
identified via photon conversions ($\gamma \to e^{+}e^{-}$), using 
various invariant mass ($m_{ee}$) cuts to account for purity-related 
systematics. The uncertainty for the 1/$\beta$ cut efficiency is determined by comparing efficiencies from the bin counting method and from fitting the 1/$\beta$ distributions with a Gaussian function. For the BEMC matching and $E_0/p$ cut efficiencies, the uncertainty is calculated as the difference between detector simulation and a data driven method based on a photonic electron sample. The overall electron identification efficiency uncertainty is estimated by adding the individual components in quadrature.

\section{Results and discussion}
\label{Results and discussion}
\par
\begin{figure}[h]
    \centering
    {\includegraphics[width=1\linewidth]{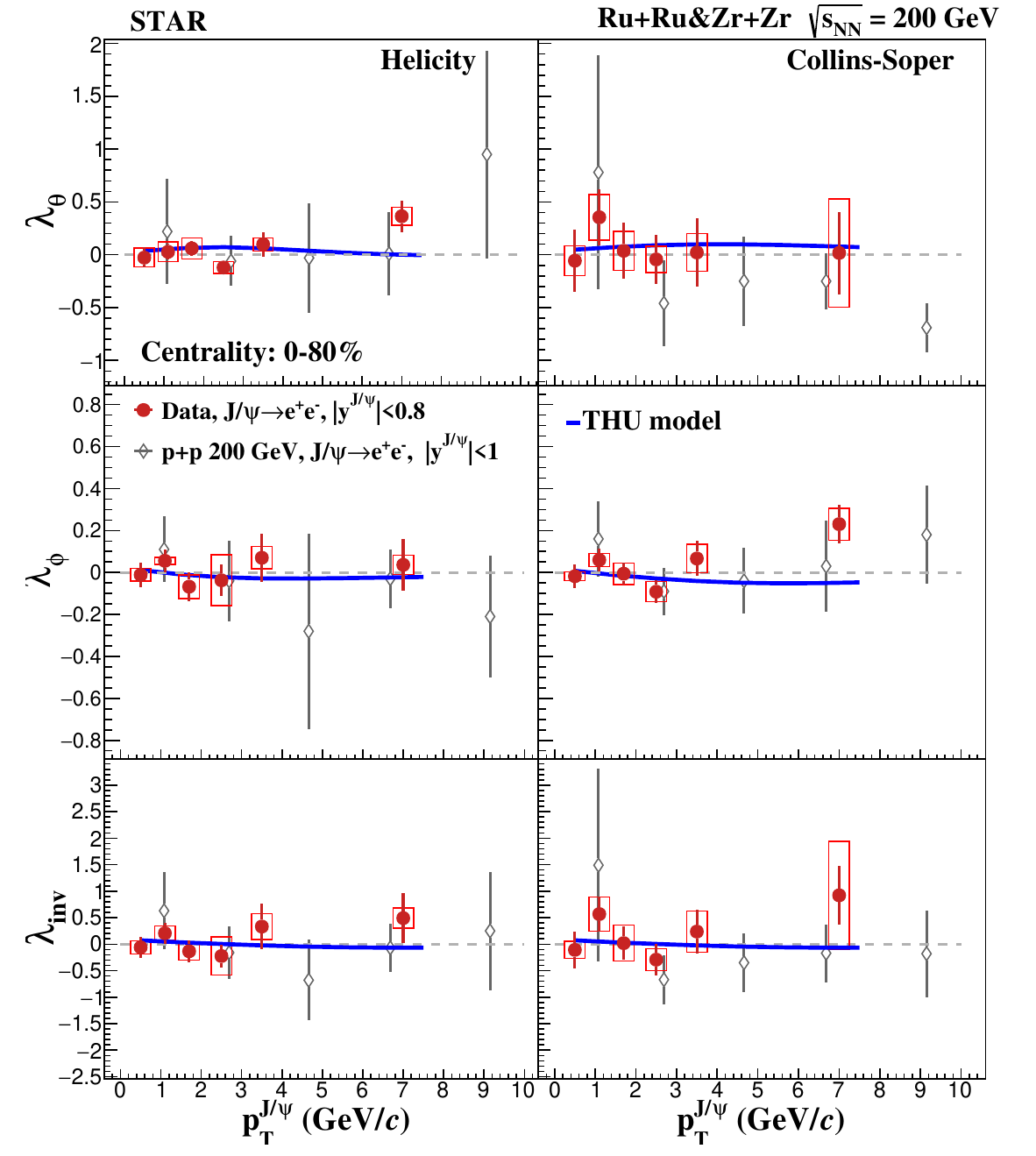}}
    \caption{Inclusive \ensuremath{J/\psi} polarization parameters (from top to bottom: $\lambda_{\theta},\lambda_{\phi},\lambda_\mathrm{inv}$) as a function of \ensuremath{J/\psi} $p_{\rm T}$ for Ru+Ru and Zr+Zr collisions at $\sqrt{s_{_{\rm NN}}} = 200$ GeV. The bars indicate statistical uncertainties, while the boxes denote systematic uncertainties. Polarization parameters in the helicity frame are presented on the left and those from Collins-Soper frame are shown on the right.}
    \label{fig:PolarizationResultpt}
    \end{figure}

 Figure \ref{fig:PolarizationResultpt} presents the inclusive \ensuremath{J/\psi} polarization parameters ($\lambda_{\theta}$, $\lambda_{\phi}$, and $\lambda_\mathrm{inv}$) as a function of $p_{\rm T}^{J/\psi}$ in 0–80\% centrality Ru+Ru and Zr+Zr collisions at $\sqrt{s_{_{\rm NN}}} = 200$ GeV. These parameters are also listed in Table~\ref{table_results pt}. In both the HX and CS frames, they are consistent with zero within uncertainties, and no significant dependence on $p_{\rm T}^{J/\psi}$ is observed.  The larger uncertainties in the CS frame arise from the limited detector acceptance in this frame, which leads to less constrained fits; these effects are further propagated to $\lambda_\mathrm{inv}$. Nevertheless, the $\lambda_\mathrm{inv}$ values shown in the bottom panels of Fig.~\ref{fig:PolarizationResultpt} are in agreement between the two frames, demonstrating the self-consistency of the results. These results are also seen to be consistent with similar measurements in $p$+$p$ collisions at $\sqrt{s}=200$ GeV within uncertainties\,\cite{STARLiuZhen2020polarization}, which are also shown in Fig.~\ref{fig:PolarizationResultpt}. It has recently been observed that $\psi(2S)$ is more suppressed than \jpsi\ in Ru+Ru and Zr+Zr collisions relative to $p$+$p$ collisions by more than a factor of 2\,\cite{psi2s_results}. This implies that $\chi_{cJ}$ could also be suppressed to a larger extent than \jpsi\ due to their smaller binding energies. Consequently, the resulting modifications to the different feed-down contributions to the inclusive \jpsi\ sample could induce variations of the inclusive \jpsi\ polarization in Ru+Ru and Zr+Zr collisions compared to those in $p$+$p$ collisions. However, the current experimental precision is insufficient to tease out such potential differences, and more precise measurements in both $p$+$p$ and heavy-ion collisions are called for.

\par
\begin{figure}[h]
    \centering
    {\includegraphics[width=1\linewidth]{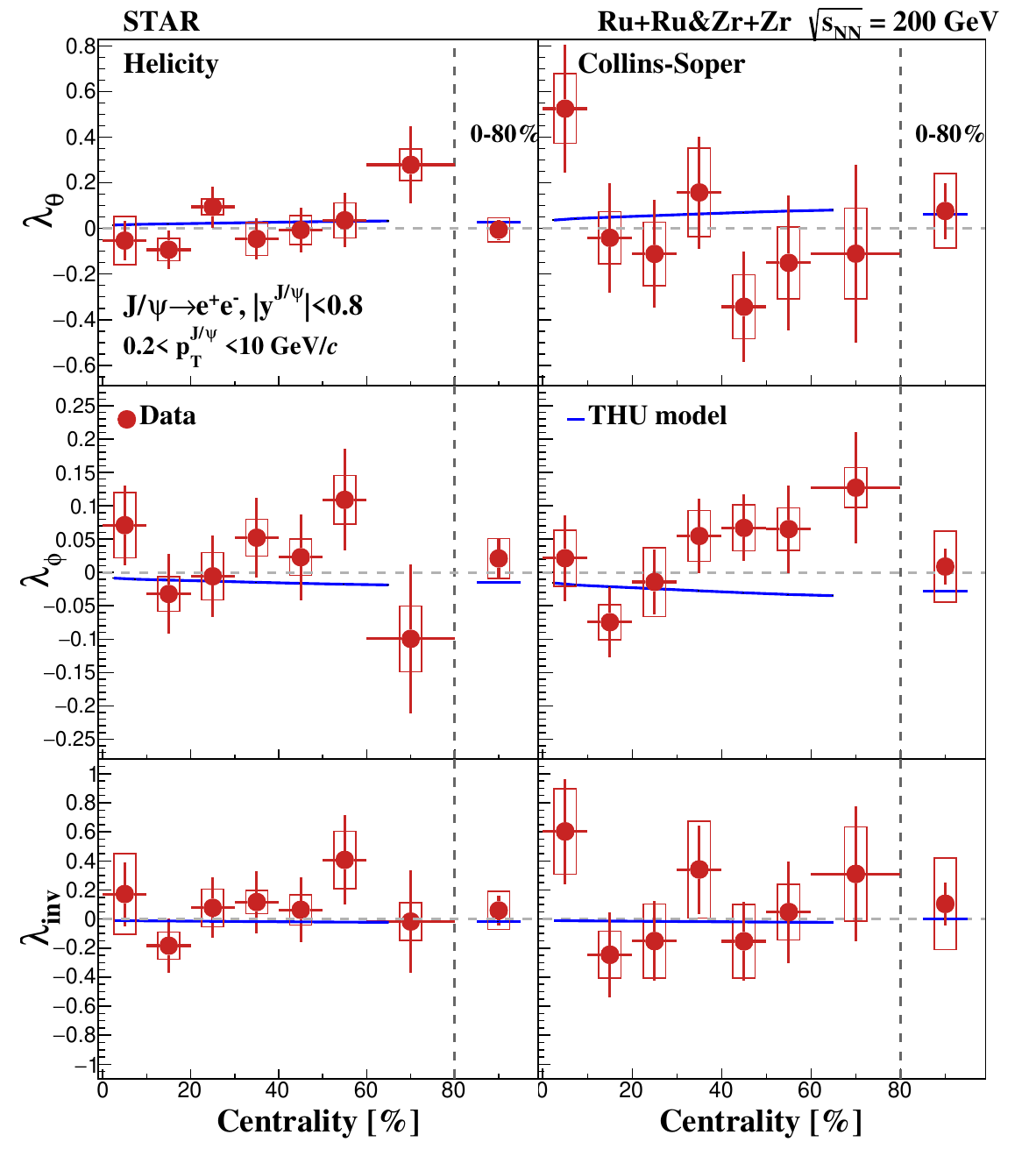}}
    \caption{The inclusive \ensuremath{J/\psi} polarization parameters (from top to bottom: $\lambda_{\theta},\lambda_{\phi},\lambda_\mathrm{inv}$) as a function of centrality, with the centrality integrated results shown in the right panel, for Ru+Ru and Zr+Zr collisions at $\sqrt{s_{_{\rm NN}}} = 200$ GeV. The bars indicate statistical uncertainties, while the boxes denote systematic uncertainties. Polarization parameters in the helicity frame are presented on the left and those from Collins-Soper frame are presented on the right.}
    \label{fig:PolarizationResultCentrality}
    \end{figure}

 The solid curves in the Figs.\,\ref{fig:PolarizationResultpt} and \ref{fig:PolarizationResultCentrality} represent the predictions of prompt \ensuremath{J/\psi} polarization from the Tsinghua (THU) model~\cite{TheroyBaochi2024polarization}. The THU model uses a relativistic Boltzmann transport equation to describe the evolution of charmonia in heavy-ion collisions, accounting for both their dissociation and regeneration. In this model, the primordial \ensuremath{J/\psi} polarization is calculated within the framework of non-relativistic quantum chromodynamics~\cite{NRQCD1, NRQCD2, NRQCD3, NRQCD4, NRQCD5}, while regenerated $J/\psi$ are assumed to be unpolarized. 
The fraction of regenerated $J/\psi$ in the model decreases 
from approximately 50\% in central collisions to less than 5\% 
in peripheral collisions~\cite{Jpsi_Regeneration:2024}. Furthermore, 
this regeneration contribution is most significant at low transverse 
momentum and decreases rapidly with increasing $p_{\text{T}}$. As mentioned previously, the contribution of non-prompt \ensuremath{J/\psi} in the inclusive \ensuremath{J/\psi} sample is estimated to be less than 15\%, depending on \ensuremath{J/\psi} $p_{\rm T}$\,\cite{Non_prompt_STAR}. The model calculations are in good agreement with experimental data.

\par Figure~\ref{fig:PolarizationResultCentrality} illustrates the dependence of the inclusive \ensuremath{J/\psi} polarization parameters within $0.2 < p_{\rm T}^{J/\psi} < 10$ GeV/$c$ on collision centrality in the HX and CS frames for Ru+Ru and Zr+Zr collisions at $\sqrt{s_{\rm NN}}=200$ GeV. The numerical values can be found in Table ~\ref{table_results cent}. As seen in Fig.~\ref{fig:PolarizationResultCentrality}, the polarization parameters are consistent with zero with no significant centrality dependence. In the analysis, the average $p_{\rm T}^{J/\psi}$ is approximately 3 GeV/$c$ and the contribution of non-prompt \ensuremath{J/\psi} is less than 5\%\,\cite{Non_prompt_STAR}. The THU model calculations for prompt \ensuremath{J/\psi} also describe the measured polarization parameters as a function of centrality reasonably well. As expected, the $\lambda_\mathrm{inv}$ values are consistent between the two frames.
    \begin{figure}[h]
    \centering
    {\includegraphics[width=1\linewidth]{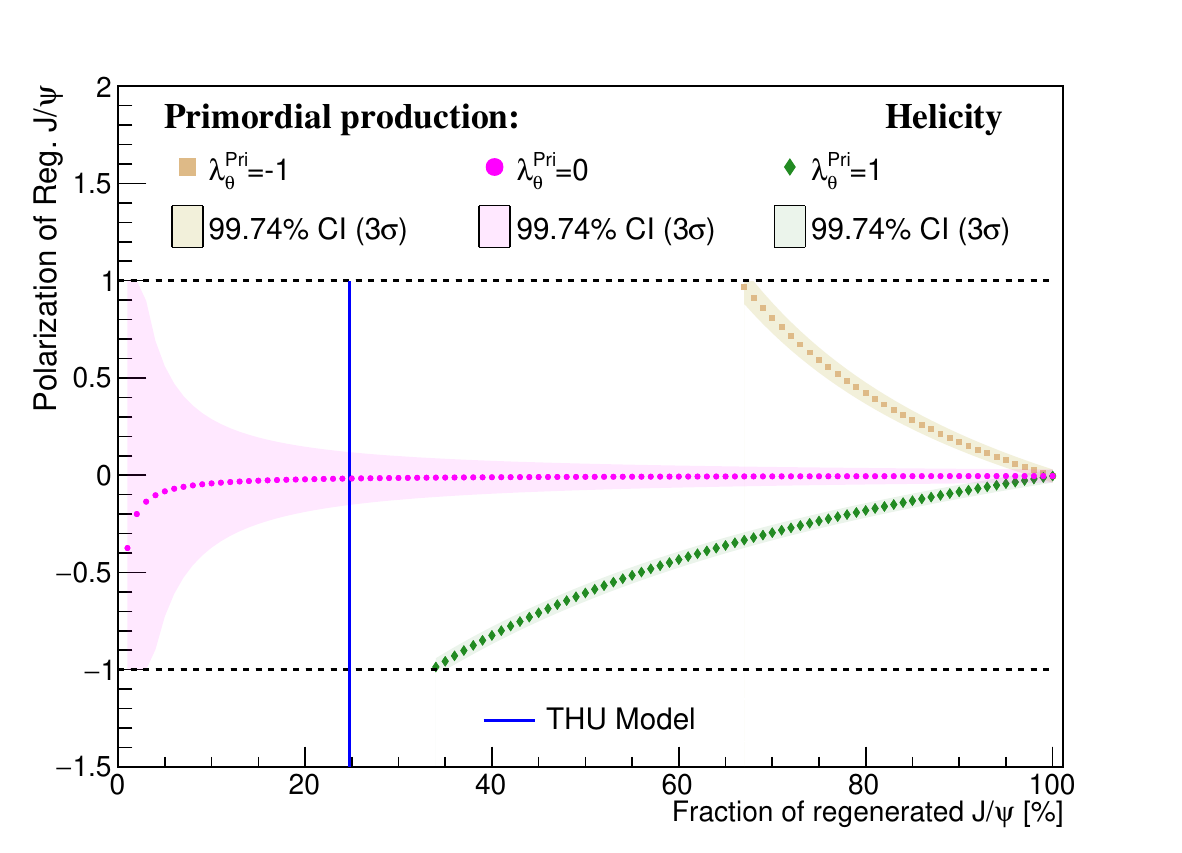}}
    \caption{Extracted polarization parameter for regenerated \ensuremath{J/\psi} in the HX frame as a function of the regenerated \ensuremath{J/\psi} fraction in the total \ensuremath{J/\psi} sample. Three different polarization values for primordial \ensuremath{J/\psi} are assumed. The bands indicate 99.74\% confidence intervals.}
    \label{fig:Reg Jpsi}
    \end{figure}
\par To shed light on the polarization of regenerated \ensuremath{J/\psi}, which likely differs from that of primordial \ensuremath{J/\psi}, the $\cos\theta$ distribution measured in the HX frame for $0.2 < p_{\rm T}^{J/\psi} < 10$ GeV/$c$ and 0–80\% centrality is fit with the following formula: 
\begin{equation}
 \begin{aligned}
      \emph W(\rm cos\theta)=3\times[(1-&P_{\rm Reg})\times\frac{1+\lambda^{\rm Pri}_{\theta}\ \rm cos^2\theta}{2\times(3+\lambda^{\rm Pri}_{\theta})}\\
      +&P_{\rm Reg}\times\frac{1+\lambda^{\rm Reg}_{\theta}\ \rm cos^2\theta}{2\times(3+\lambda^{\rm Reg}_{\theta})}],
    \label{Reg.jpsi Lambda_theta}
  \end{aligned}
\end{equation}
where $P_{\text{Reg}}$ represents the fraction of regenerated \ensuremath{J/\psi}, and $\lambda_{\theta}^{\text{Pri}}$ and $\lambda_{\theta}^{\text{Reg}}$ are the polarization parameters for primordial and regenerated \ensuremath{J/\psi}, respectively. $P_{\text{Reg}}$ is scanned over the range of 0 to 1, while three special values, i.e., 1, 0, and -1, are assumed for $\lambda_{\theta}^{\text{Pri}}$. 
The resulting regenerated \ensuremath{J/\psi} $\lambda_{\theta}$ in the HX frame is shown in Fig.~\ref{Reg.jpsi Lambda_theta} as a function of the regenerated \ensuremath{J/\psi} fraction in the inclusive \ensuremath{J/\psi} sample. The vertical dashed line at about 25\% indicates the fraction of regenerated \ensuremath{J/\psi} predicted by the THU model. Different shaded bands correspond to different assumed $\lambda_{\theta}^{\text{Pri}}$ values. When the primordial \ensuremath{J/\psi} is unpolarized, the regenerated \ensuremath{J/\psi} also tends to have zero polarization. This exercise provides a basis for future extraction of the polarization of regenerated \ensuremath{J/\psi} once the knowledge of the primordial \ensuremath{J/\psi} polarization and the regenerated \ensuremath{J/\psi} fraction is improved.

\begin{table*}[!ht]
\renewcommand{\arraystretch}{1.4}
\setlength{\tabcolsep}{20pt}
\centering
\caption{Inclusive \ensuremath{J/\psi} polarization parameters in the HX and CS frames for different $p_{\rm T}^{J/\psi}$ bins within $|y^{J/\psi}|<0.8$ in 0–80\% Ru+Ru and Zr+Zr collisions at $\sqrt{s_{\rm {NN}}} = 200$~GeV. The first uncertainty is statistical and the second systematic.}
\label{tab:systopo}
\begin{tabular}{l |c|c| c}
\hline\hline
 $p_{\rm T}$ (GeV/$c$) & $\lambda_{\theta}^{\rm HX}$ & $\lambda_{\phi}^{\rm HX}$ & $\lambda_\mathrm{inv}^{\rm HX}$\\
 \hline
[0.2,0.8) & -0.027 $\pm$ 0.095 $\pm$ 0.089 & -0.011 $\pm$ 0.059 $\pm$ 0.031 & -0.060 $\pm$ 0.197 $\pm$ 0.123 \\ 
\hline
[0.8,1.4) & 0.028 $\pm$ 0.074 $\pm$ 0.092 & 0.056 $\pm$ 0.054 $\pm$ 0.016 & 0.206 $\pm$ 0.197 $\pm$ 0.136 \\ 
\hline
[1.4,2.0) & 0.060 $\pm$ 0.074 $\pm$ 0.099 & -0.068 $\pm$ 0.070 $\pm$ 0.056 & -0.133 $\pm$ 0.203 $\pm$ 0.165 \\ 
\hline
[2.0,3.0) & -0.121 $\pm$ 0.061 $\pm$ 0.056 & -0.037 $\pm$ 0.076 $\pm$ 0.121 & -0.224 $\pm$ 0.213 $\pm$ 0.357 \\ 
\hline
[3.0,4.0) & 0.099 $\pm$ 0.116 $\pm$ 0.062 & 0.070 $\pm$ 0.116 $\pm$ 0.053 & 0.332 $\pm$ 0.430 $\pm$ 0.242 \\ 
\hline
[4.0,10.0) & 0.364 $\pm$ 0.149 $\pm$ 0.085 & 0.036 $\pm$ 0.124 $\pm$ 0.046 & 0.492 $\pm$ 0.472 $\pm$ 0.192 \\ 
\hline
 $p_{\rm T}$ (GeV/$c$) & $\lambda_{\theta}^{\rm CS}$ & $\lambda_{\phi}^{\rm CS}$ & $\lambda_\mathrm{inv}^{\rm CS}$\\
 \hline
[0.2,0.8) & -0.056 $\pm$ 0.298 $\pm$ 0.141 & -0.018 $\pm$ 0.056 $\pm$ 0.021 & -0.108 $\pm$ 0.343 $\pm$ 0.161 \\ 
\hline
[0.8,1.4) & 0.355 $\pm$ 0.269 $\pm$ 0.217 & 0.060 $\pm$ 0.055 $\pm$ 0.032 & 0.568 $\pm$ 0.327 $\pm$ 0.325 \\ 
\hline
[1.4,2.0) & 0.037 $\pm$ 0.263 $\pm$ 0.185 & -0.006 $\pm$ 0.053 $\pm$ 0.051 & 0.020 $\pm$ 0.309 $\pm$ 0.338 \\ 
\hline
[2.0,3.0) & -0.044 $\pm$ 0.236 $\pm$ 0.122 & -0.092 $\pm$ 0.055 $\pm$ 0.048 & -0.292 $\pm$ 0.291 $\pm$ 0.210 \\ 
\hline
[3.0,4.0) & 0.022 $\pm$ 0.321 $\pm$ 0.177 & 0.067 $\pm$ 0.084 $\pm$ 0.067 & 0.237 $\pm$ 0.416 $\pm$ 0.386 \\ 
\hline
[4.0,10.0) & 0.017 $\pm$ 0.390 $\pm$ 0.512 & 0.231 $\pm$ 0.093 $\pm$ 0.076 & 0.921 $\pm$ 0.560 $\pm$ 1.021 \\ 
\hline
\end{tabular}
\label{table_results pt}
\end{table*}

\begin{table*}[!ht]
\renewcommand{\arraystretch}{1.4}
\setlength{\tabcolsep}{20pt}
\centering
\caption{Inclusive \ensuremath{J/\psi} polarization parameters, integrated over $0.2<p_{\rm T}^{J/\psi}<10$ GeV/$c$ and within $|y^{J/\psi}|<0.8$, in the HX and CS frames in different centrality bins of Ru+Ru and Zr+Zr collisions at $\sqrt{s_{\rm {NN}}} = 200$ GeV. The first uncertainty is statistical and the second systematic.}
\label{tab:systopo}
\begin{tabular}{l |c|c| c}
\hline\hline
 Centrality [\%] & $\lambda_{\theta}^{\rm HX}$ & $\lambda_{\phi}^{\rm HX}$ & $\lambda_\mathrm{inv}^{\rm HX}$\\
 \hline
[0,10] & -0.053 $\pm$ 0.087 $\pm$ 0.105 & 0.071 $\pm$ 0.060 $\pm$ 0.049 & 0.172 $\pm$ 0.220 $\pm$ 0.277 \\ 
\hline
[10,20] & -0.093 $\pm$ 0.085 $\pm$ 0.049 & -0.032 $\pm$ 0.060 $\pm$ 0.026 & -0.184 $\pm$ 0.185 $\pm$ 0.094 \\ 
\hline
[20,30] & 0.095 $\pm$ 0.090 $\pm$ 0.035 & -0.006 $\pm$ 0.061 $\pm$ 0.035 & 0.077 $\pm$ 0.207 $\pm$ 0.130 \\ 
\hline
[30,40] & -0.046 $\pm$ 0.089 $\pm$ 0.073 & 0.052 $\pm$ 0.059 $\pm$ 0.028 & 0.117 $\pm$ 0.213 $\pm$ 0.079 \\ 
\hline
[40,50] & -0.007 $\pm$ 0.099 $\pm$ 0.063 & 0.023 $\pm$ 0.064 $\pm$ 0.027 & 0.063 $\pm$ 0.224 $\pm$ 0.104 \\ 
\hline
[50,60] & 0.036 $\pm$ 0.119 $\pm$ 0.077 & 0.109 $\pm$ 0.076 $\pm$ 0.037 & 0.407 $\pm$ 0.310 $\pm$ 0.198 \\ 
\hline
[60,80] & 0.279 $\pm$ 0.170 $\pm$ 0.069 & -0.099 $\pm$ 0.112 $\pm$ 0.049 & -0.017 $\pm$ 0.356 $\pm$ 0.130 \\ 
\hline
[0,80] & -0.005 $\pm$ 0.043 $\pm$ 0.053 & 0.021 $\pm$ 0.030 $\pm$ 0.030 & 0.059 $\pm$ 0.103 $\pm$ 0.132 \\ 
\hline
 Centrality [\%] & $\lambda_{\theta}^{\rm CS}$ & $\lambda_{\phi}^{\rm CS}$ & $\lambda_\mathrm{inv}^{\rm CS}$\\
 \hline
[0,10] & 0.526 $\pm$ 0.279 $\pm$ 0.153 & 0.021 $\pm$ 0.065 $\pm$ 0.042 & 0.603 $\pm$ 0.362 $\pm$ 0.293 \\ 
\hline
[10,20] & -0.041 $\pm$ 0.241 $\pm$ 0.115 & -0.075 $\pm$ 0.052 $\pm$ 0.026 & -0.246 $\pm$ 0.291 $\pm$ 0.162 \\ 
\hline
[20,30] & -0.112 $\pm$ 0.235 $\pm$ 0.140 & -0.014 $\pm$ 0.049 $\pm$ 0.052 & -0.152 $\pm$ 0.275 $\pm$ 0.255 \\ 
\hline
[30,40] & 0.158 $\pm$ 0.246 $\pm$ 0.194 & 0.055 $\pm$ 0.056 $\pm$ 0.038 & 0.341 $\pm$ 0.305 $\pm$ 0.334 \\ 
\hline
[40,50] & -0.344 $\pm$ 0.243 $\pm$ 0.141 & 0.067 $\pm$ 0.050 $\pm$ 0.035 & -0.153 $\pm$ 0.273 $\pm$ 0.254 \\ 
\hline
[50,60] & -0.151 $\pm$ 0.296 $\pm$ 0.158 & 0.065 $\pm$ 0.066 $\pm$ 0.032 & 0.047 $\pm$ 0.350 $\pm$ 0.193 \\ 
\hline
[60,80] & -0.111 $\pm$ 0.391 $\pm$ 0.199 & 0.127 $\pm$ 0.084 $\pm$ 0.030 & 0.310 $\pm$ 0.467 $\pm$ 0.324 \\ 
\hline
[0,80] & 0.077 $\pm$ 0.123 $\pm$ 0.164 & 0.009 $\pm$ 0.027 $\pm$ 0.053 & 0.104 $\pm$ 0.148 $\pm$ 0.316 \\ 
\hline
\end{tabular}
\label{table_results cent}
\end{table*}

\section{Summary}
\label{summary}
 The STAR experiment at RHIC presents the first measurements of inclusive \ensuremath{J/\psi} polarization in the helicity and Collins-Soper frames in Ru+Ru and Zr+Zr collisions at $\sqrt{s_{_{\rm NN}}} = 200$~GeV. The polarization parameters ($\lambda_{\theta}\rm$, $\lambda_{\phi}$, and $\lambda_\mathrm{inv}$) are studied as a function of $p_{T}^{J/\psi}$ and collision centrality. They are found to be consistent with zero in the $p^{J/\psi}_{\rm T}$ range of 0.2 to 10 GeV/$c$ and a centrality range of 0–80\%. They are also found to be consistent with similar measurements in 200 GeV $p$+$p$ collisions, and can be well described by a transport model calculation for prompt \ensuremath{J/\psi}, in which regenerated \ensuremath{J/\psi} are assumed to be unpolarized. These results provide further insights into \ensuremath{J/\psi} production and propagation in the QGP, which in turn will help improve our understanding of QGP properties.

\section*{Acknowledgements}
We thank the RHIC Operations Group and SDCC at BNL, the NERSC Center at LBNL, and the Open Science Grid consortium for providing resources and support.  This work was supported in part by the Office of Nuclear Physics within the U.S. DOE Office of Science, the U.S. National Science Foundation, National Natural Science Foundation of China, Chinese Academy of Science, the Ministry of Science and Technology of China and the Chinese Ministry of Education, NSTC Taipei, the National Research Foundation of Korea, Czech Science Foundation and Ministry of Education, Youth and Sports of the Czech Republic, Hungarian National Research, Development and Innovation Office, New National Excellency Programme of the Hungarian Ministry of Human Capacities, Department of Atomic Energy and Department of Science and Technology of the Government of India, the National Science Centre and WUT ID-UB of Poland, German Bundesministerium f\"ur Bildung, Wissenschaft, Forschung and Technologie (BMBF), Helmholtz Association, Ministry of Education, Culture, Sports, Science, and Technology (MEXT), and Japan Society for the Promotion of Science (JSPS).

\appendix



\bibliographystyle{unsrt} 
\bibliography{reference.bib}

@article{jpsi_suppresion_in_QGP1986j,
  author = "Matsui, T. and Satz, H.",
  title = "{J/$\psi$ Suppression by Quark-Gluon Plasma Formation}",
  doi = "10.1016/0370-2693(86)91404-8",
  journal = "Phys. Lett. B",
  volume = "178",
  pages = "416--422",
  year = "1986"
}

@article{enhanced_jpsi_qgp_2001enhanced,
  author = "Thews, R. L. and Schroedter, M. and Rafelski, J.",
  title = "{Enhanced J/$\psi$ Production in Deconfined Quark Matter}",
  doi = "10.1103/PhysRevC.63.054905",
  journal = "Phys. Rev. C",
  volume = "63",
  pages = "054905",
  year = "2001"
}

@article{Braun-Munzinger:2015hba,
  author  = {Braun-Munzinger, P. and Koch, V. and Sch{\"a}fer, T. and Stachel, J.},
  title   = {Properties of hot and dense matter from relativistic heavy-ion collisions},
  journal = {Phys. Rep.},
  volume  = {621},
  pages   = {76--126},
  year    = {2016}
}

@article{Harris:2023tti,
  author  = {Harris, J. W. and M{\"u}ller, B.},
  title   = {{QGP} signatures revisited},
  journal = {Eur. Phys. J. C},
  volume  = {84},
  number  = {3},
  pages   = {247},
  year    = {2024}
}

@article{Andronic:2025jbp,
    author  = {A. Andronic and R. Arnaldi and others},
    title   = "{Quarkonia and Deconfined Quark-Gluon Matter in Heavy-Ion Collisions}",
    journal = {arXiv preprint arXiv:2501.08290},
    year    = {2025}
}

@article{Tang:2020ame,
  author = "Tang, Z. B. and Zha, W. M. and Zhang, Y. F.",
  title = "{An experimental review of open heavy flavor and quarkonium production at RHIC}",
  eprint = "2105.11656",
  archivePrefix = "arXiv",
  primaryClass = "nucl-ex",
  doi = "10.1007/s41365-020-00785-8",
  journal = "Nucl. Sci. Tech.",
  volume = "31",
  number = "8",
  pages = "81",
  year = "2020"
}

@article{Rothkopf:2019ipj,
  author = "Rothkopf, A.",
  title = "{Heavy Quarkonium in Extreme Conditions}",
  eprint = "1912.02253",
  archivePrefix = "arXiv",
  primaryClass = "hep-ph",
  doi = "10.1016/j.physrep.2020.02.006",
  journal = "Phys. Rept.",
  volume = "858",
  pages = "1--117",
  year = "2020"
}

@article{Theroyfaccioli2010towardsIntroduction,
  author = "Faccioli, P. and others",
  title = "{Towards the Experimental Clarification of Quarkonium Polarization}",
  doi = "10.1140/epjc/s10052-010-1368-9",
  journal = "Eur. Phys. J. C",
  volume = "69",
  pages = "657--673",
  year = "2010"
}

@book{BookingParticlefaccioli2023polarization,
  author = "Faccioli, P. and Louren{\c{c}}o, C.",
  title = "{Particle Polarization in High Energy Physics: An Introduction and Case Studies on Vector Particle Production at the LHC}",
  publisher = "Springer Nature",
  year = "2023"
}

@article{TheroyBLioffe2003quarkonium,
  author = "Ioffe, B. L. and Kharzeev, D. E.",
  title = "{Quarkonium Polarization in Heavy Ion Collisions as a Possible Signature of the Quark-Gluon Plasma}",
  doi = "10.1103/PhysRevC.68.061902",
  journal = "Phys. Rev. C",
  volume = "68",
  pages = "061902",
  year = "2003"
}

@article{TheroyBaochi2024polarization,
  author = "Zhao, J. and Shi, S.",
  title = "{Detecting Nuclear Mass Distribution in Isobar Collisions via Charmonium}",
  doi = "10.1140/epjc/s10052-023-11827-5",
  journal = "Eur. Phys. J. C",
  volume = "83",
  pages = "571",
  year = "2023"
}

@article{STARLiuZhen2020polarization,
  author = "Adam, J. and others",
  collaboration = "STAR",
  title = "{Measurement of inclusive J/$\psi$ polarization in $p$+$p$ collisions at $\sqrt {s}$ = 200 GeV by the STAR experiment}",
  eprint = "2007.04732",
  archivePrefix = "arXiv",
  primaryClass = "hep-ex",
  doi = "10.1103/PhysRevD.102.092009",
  journal = "Phys. Rev. D",
  volume = "102",
  number = "9",
  pages = "092009",
  year = "2020"
}

@article{LHCHIC2021firstpolarizarion,
  author = "Acharya, S. and others",
  collaboration = "ALICE",
  title = "{First measurement of quarkonium polarization in nuclear collisions at the LHC}",
  doi = "10.1016/j.physletb.2021.136146",
  journal = "Phys. Lett. B",
  volume = "815",
  pages = "136146",
  year = "2021"
}

@article{LHCLHCb2013polarizationpp,
  author = "Aaij, R. and others",
  collaboration = "LHCb",
  title = "{Measurement of J/$\psi$ polarization in $pp$ collisions at $\sqrt{s}$ = 7 TeV}",
  doi = "10.1140/epjc/s10052-013-2431-9",
  journal = "Eur. Phys. J. C",
  volume = "73",
  pages = "2431",
  year = "2013"
}

@article{Non_prompt_STAR,
  author = "Adamczyk, L. and others",
  collaboration = "STAR",
  title = "{$\rm {J}/\psi$ production at high transverse momenta in $p+p$ and Au+Au collisions at $\sqrt{s_{NN}}$ = 200 GeV}",
  eprint = "1208.2736",
  archivePrefix = "arXiv",
  primaryClass = "nucl-ex",
  doi = "10.1016/j.physletb.2013.04.010",
  journal = "Phys. Lett. B",
  volume = "722",
  pages = "55--62",
  year = "2013"
}

@article{AuAu200_2019measurement,
  author = "Adam, J. and others",
  collaboration = "STAR",
  title = "{Measurement of inclusive J/$\psi$ suppression in Au+Au collisions at $\sqrt{s_{NN}}$ = 200 GeV through the dimuon channel at STAR}",
  doi = "10.1016/j.physletb.2019.134917",
  journal = "Phys. Lett. B",
  volume = "797",
  pages = "134917",
  year = "2019"
}

@article{STARTPC2003star,
  title = "{The STAR time projection chamber: a unique tool for studying high multiplicity events at RHIC}",
  author = "Anderson, M. and others",
  journal = "Nucl. Instrum. Methods Phys. Res. A",
  volume = "499",
  number = "2-3",
  pages = "659--678",
  year = "2003",
  publisher = "Elsevier"
}

@article{STARmagneticfield2003star,
  title = "{The STAR detector magnet subsystem}",
  author = "Bergsma, F. and others",
  journal = "Nucl. Instrum. Methods Phys. Res. A",
  volume = "499",
  number = "2-3",
  pages = "633--639",
  year = "2003",
  publisher = "Elsevier"
}

@article{STARBEMC2003star,
  title = "{The STAR barrel electromagnetic calorimeter}",
  author = "Beddo, M. and others",
  journal = "Nucl. Instrum. Methods Phys. Res. A",
  volume = "499",
  number = "2-3",
  pages = "725--739",
  year = "2003",
  publisher = "Elsevier"
}

@article{TOF2003single,
  title = "{A single Time-of-Flight tray based on multigap resistive plate chambers for the STAR experiment at RHIC}",
  author = "Bonner, B. and others",
  journal = "Nucl. Instrum. Methods Phys. Res. A",
  volume = "508",
  number = "1-2",
  pages = "181--184",
  year = "2003",
  publisher = "Elsevier"
}

@article{Xu:2016alq,
  author = "Xu, Y. F. and others",
  title = "{Physics performance of the STAR zero degree calorimeter at relativistic heavy ion collider}",
  doi = "10.1007/s41365-016-0129-z",
  journal = "Nucl. Sci. Tech.",
  volume = "27",
  number = "6",
  pages = "126",
  year = "2016"
}

@article{photonicConversiondetailed,
  title = "{Detailed measurement of the e$^+$ e$^-$ pair continuum in p+p and Au+Au collisions at $\sqrt{s_{NN}}=200$ GeV and implications for direct photon production}",
  author = "Adare, A. and others",
  journal = "Phys. Rev. C",
  volume = "81",
  number = "3",
  pages = "034911",
  year = "2010",
  publisher = "APS"
}

@techreport{Bichselfunction2001comparison,
  title = "{Comparison of Bethe-Bloch and Bichsel functions, STAR note SN0439}",
  author = "Bichsel, H.",
  year = "2001",
  institution = "Technical report, Dec. 20"
}

@article{barlow2002systematic,
  title = "{Systematic errors: facts and fictions}",
  author = "Barlow, R.",
  journal = "arXiv preprint hep-ex/0207026",
  year = "2002"
}

@article{PHENIXRapidity2009j,
  title = "{J/$\psi$ production measurements by the PHENIX experiment}",
  author = "Atomssa, E.T. and others",
  journal = "Eur. Phys. J. C",
  volume = "61",
  pages = "683--686",
  year = "2009",
  publisher = "Springer"
}

@article{NRQCD1,
  title = "{Rigorous QCD analysis of inclusive annihilation and production of heavy quarkonium}",
  author = "Bodwin, G.T. and Braaten, E. and Lepage, G.P.",
  journal = "Phys. Rev. D",
  volume = "51",
  number = "3",
  pages = "1125",
  year = "1995",
  publisher = "APS"
}

@article{NRQCD2,
  title = "{J/$\psi$ polarization at the Tevatron and the LHC: nonrelativistic-QCD factorization at the crossroads}",
  author = "Butenschoen, M. and Kniehl, B.A.",
  journal = "Phys. Rev. D",
  volume = "82",
  number = "3",
  pages = "034029",
  year = "2010",
  publisher = "APS"
}

@article{NRQCD3,
  title = "{Polarization for prompt J/$\psi$ and $\psi$(2s) production at the Tevatron and LHC}",
  author = "Gong, B. and Wan, L.P. and Wang, J.X. and Zhang, H.F.",
  journal = "Phys. Rev. Lett.",
  volume = "110",
  year = "2013",
  publisher = "APS"
}

@article{NRQCD4,
  title = "{J/$\psi$ polarization at hadron colliders in nonrelativistic QCD}",
  author = "Chao, K. T. and others",
  journal = "Phys. Rev. Lett.",
  volume = "108",
  number = "24",
  pages = "242004",
  year = "2012",
  publisher = "APS"
}

@article{NRQCD5,
  title = "{J/$\psi$ polarization in the CGC+ NRQCD approach}",
  author = "Ma, Y. Q. and others",
  journal = "J. High Energy Phys.",
  volume = "2018",
  number = "12",
  pages = "1--27",
  year = "2018",
  publisher = "Springer"
}

@article{mix_event_STAR_phi_2004,
  title = "{$\Phi$ meson production in Au+Au and p+p collisions at $\sqrt{s_{NN}}=200$ GeV}",
  author = "Adam, J. and others",
  collaboration = "STAR",
  journal = "Phys. Rev. C",
  volume = "71",
  number = "6",
  pages = "064902",
  year = "2005",
  publisher = "APS",
  doi = "10.1103/PhysRevC.71.064902"
}

@article{photon_fully_polarization_2011determination,
    author  = {P. Faccioli and C. Louren\c{c}o and J. Seixas and H. K. W\"{o}hri},
    title   = "{Determination of $\chi_c$ and $\psi'$ polarization in high-energy collisions}",
    journal = {Phys. Rev. D},
    volume  = {83},
    year    = {2011},
    pages   = {096001},
    doi     = {10.1103/PhysRevD.83.096001}
}

@article{isobarCentrality2022search,
  title = "{Search for the chiral magnetic effect with isobar collisions at $\sqrt{s_{NN}}$ = 200 GeV by the STAR Collaboration at the BNL Relativistic Heavy Ion Collider}",
  author = "Aboona, BE and Adams, JR and Adkins, JK and Aggarwal, MM and others",
  journal = "Phys. Rev. C",
  volume = "105",
  number = "1",
  pages = "014901",
  year = "2022",
  publisher = "APS"
}

@article{crystal_ball_1983charmonium,
  title = "{Charmonium spectroscopy from radiative decays of the $\rm{J}/\psi$ and $\psi'$}",
  author = "Gaiser, John Erthal",
  journal = "Stanford University",
  year = "1983"
}

@article{crystal_ball_1986study,
  title = "{A study of the radiative cascade transitions between the $\Upsilon'$ and $\Upsilon$ resonances}",
  author = "Skwarnicki, Tomasz",
  journal = "DESY",
  year = "1986"
}

@article{crystal_ball_1982study,
  title = "{Study of the reaction $\psi^{'} \to \gamma\gamma\psi$}",
  author = "M. Oreglia and D. Butler and ... and N. AuthorLast",
  journal = "Phys. Rev. D",
  volume = "25",
  number = "9",
  pages = "2259",
  year = "1982",
  publisher = "APS"
}

@misc{ROOT:RooCrystalBall,
    author = "{ROOT Collaboration}",
    title = "{RooCrystalBall Class Reference}",
    year = "2025",
    howpublished = "\url{https://root.cern.ch/doc/master/classRooCrystalBall.html}",
    note = "[Online; accessed 1-September-2025]"
}

@article{Jpsi_Regeneration:2024,
    author = "Zhao, Jiaxing and Chen, Baoyi",
    title = "{$J/\psi $ polarization in relativistic heavy ion collisions}",
    eprint = "2312.01799",
    archivePrefix = "arXiv",
    primaryClass = "hep-ph",
    doi = "10.1140/epjc/s10052-024-13264-w",
    journal = "Eur. Phys. J. C",
    volume = "84",
    number = "8",
    pages = "875",
    year = "2024"
}

@article{psi2s_results,
  author       = {B. E. Aboona and others},
  collaboration= {STAR Collaboration},
  title        = {Observation of charmonium sequential suppression in heavy-ion collisions at the Relativistic Heavy Ion Collider},
  journal      = {Phys. Rev. Lett.},
  volume       = {136},
  pages        = {122302},
  year         = {2026},
  doi          = {10.1103/ghzh-kv8z},
  note         = {Published 3 March 2026; full author list available at arXiv:2509.12842}
}

@article{psi2s_no_prolarization,
  title={Observation of $\chi_{c}$ and $\chi_b$ nuclear suppression via dilepton polarization measurements},
  author={Faccioli, Pietro and Seixas, Joao},
  journal={Phys. Rev. D},
  volume={85},
  number={7},
  pages={074005},
  year={2012},
  publisher={APS}
}

@article{SPS_QGP_2000,
  author  = {Heinz, Ulrich and Jacob, Maurice},
  title   = {Evidence for a new state of matter: An assessment of the results from the {CERN} lead beam programme},
  journal = {arXiv preprint nucl-th/0002042},
  year    = {2000},
  eprint  = {nucl-th/0002042},
  archivePrefix = {arXiv}
}

@article{Chen2025VectorMesonSpinAlignment,
  author  = {Chen, Jin-Hui and Liang, Zuo-Tang and Ma, Yu-Gang and Sheng, Xin-Li and Wang, Qun},
  title   = {Vector meson’s spin alignments in high energy reactions},
  journal = {Science China Physics, Mechanics \& Astronomy},
  year    = {2025},
  volume  = {68},
  number  = {2},
  pages   = {211001},
  doi     = {10.1007/s11433-024-2495-1},
  url     = {https://doi.org/10.1007/s11433-024-2495-1}
}






\end{document}